\documentclass[sigconf, authorversion]{acmart}
\AtBeginDocument{%
  \providecommand\BibTeX{{%
    \normalfont B\kern-0.5em{\scshape i\kern-0.25em b}\kern-0.8em\TeX}}}

\setcopyright{acmcopyright}
\copyrightyear{2023}
\acmYear{2023}
\acmDOI{XXXXXXX.XXXXXXX}

\acmConference[SIGIR '23]{SIGIR 2023}{}{ USA}
\acmPrice{15.00}
\acmISBN{978-1-4503-XXXX-X/18/06}

\usepackage{libertine}
\graphicspath{ {./images/} }
\usepackage{graphicx}
\usepackage{caption}
\usepackage{subcaption}
\usepackage{multirow}

\copyrightyear{2023}
\acmYear{2023}
\setcopyright{acmlicensed}\acmConference[SIGIR '23]{Proceedings of the 46th International ACM SIGIR Conference on Research and Development in Information Retrieval}{July 23--27, 2023}{Taipei, Taiwan}
\acmBooktitle{Proceedings of the 46th International ACM SIGIR Conference on Research and Development in Information Retrieval (SIGIR '23), July 23--27, 2023, Taipei, Taiwan}
\acmPrice{15.00}
\acmDOI{10.1145/3539618.3592004}
\acmISBN{978-1-4503-9408-6/23/07}

\begin{document}

\title{Inference at Scale}
\subtitle{Significance Testing for Large Search and Recommendation Experiments}

\author{Ngozi Ihemelandu}
\affiliation{%
  \institution{Boise State University}
  \streetaddress{}
  \city{Boise}
  \state{Idaho}
  \country{USA}}
   \email{ngoziihemelandu@u.boisestate.edu}

\author{Michael D. Ekstrand}
\affiliation{%
  \institution{Boise State University}
  \city{Boise}
  \state{Idaho}
  \country{USA}}
  \email{ekstrand@acm.org}


\begin{abstract}
 A number of information retrieval studies have been done to assess which statistical techniques are appropriate for comparing systems. However, these studies are focused on TREC-style experiments, which typically have fewer than 100 topics. There is no similar line of work for large search and recommendation experiments; such studies typically have thousands of topics or users and much sparser relevance judgements, so it is not clear if recommendations for analyzing traditional TREC experiments apply to these settings.
 In this paper, we empirically study the behavior of significance tests with large search and recommendation evaluation data. Our results show that the Wilcoxon and Sign tests show significantly higher Type-1 error rates for large sample sizes than the bootstrap, randomization and t-tests, which were more consistent with the expected error rate. While the statistical tests displayed differences in their power for smaller sample sizes, they showed no difference in their power for large sample sizes. We recommend the sign and Wilcoxon tests should not be used to analyze large scale evaluation results. Our result demonstrate that with Top-$N$ recommendation and large search evaluation data, most tests would have a $100\%$ chance of finding statistically significant results. Therefore, the effect size should be used to determine practical or scientific significance.
\end{abstract}

\ccsdesc[500]{Information systems~Retrieval effectiveness}
\ccsdesc[500]{Information systems~Retrieval efficiency}
\ccsdesc[300]{Information systems~Relevance assessment}

\ccsdesc[500]{Information systems~Recommender systems}

\keywords{evaluation, statistical inference}

\maketitle

\section{Introduction}
A key goal in information retrieval (IR) and related research is to make progress by promoting only those newly proposed algorithmic methods that truly improve the effectiveness of the state-of-the-art.  Statistical significance testing plays an important role in achieving this goal.  IR researchers use statistical significance tests to assess if an observed improvement is significant or literally due to random chance.

While there has been substantial discussion of the need for statistical significance in IR research \cite{sakai2016statistical}, recent survey of papers published at ACM RecSys found that over half of the papers examined did not use any significance test to analyze their evaluation results \citep{ihemelandu2021statistical}. One hurdle in addressing this is that there is little to no practical guidance on how to do inference for top-$N$ recommendation experiments.
There is, however, research and guidance on which statistical tests are the most appropriate to use for classic TREC-style experiments (small-scale ad-hoc search) \citep{hull1993using, sanderson2005information, zobel1998reliable, smucker2007comparison, parapar2020using, urbano2019statistical}.  Our goal is to determine if that advice holds for top-$N$ recommendation experiments, along with large search experiments that share many similar properties. 

Top-$N$ recommendation tasks and large-scale search tasks, such as MS-MARCO document or passage ranking \citep{craswell2021ms, lin2021significant} have several distinguishing features: they have many topics ($1,000$ or more vs. $50$ for a TREC track); relevance data is sparse and incomplete compared with the (approximately) complete ground-truth relevance judgements obtained by pooling in TREC experiments; and in the top-$N$ recommendation case, a few items are often relevant to many users resulting in a popularity skew, in contrast to a TREC experiment where documents are not concentrated to just a few topics.

In this paper, we empirically investigate how the pairwise significance tests --- $t$-test, bootstrap, randomization, Wilcoxon signed-rank, and sign tests --- which are commonly used to evaluate two IR systems using the same topic sets and which were studied in previous literature \citep{parapar2020using, smucker2007comparison, urbano2019statistical} behave with top-$N$ recommendation system and large search evaluation data. We employ the simulation methodology proposed by \citet{urbano2018stochastic} to empirically compare these statistical tests under the full knowledge of the null hypothesis. Our study addresses the following questions:

\begin{itemize}
    \item When two systems have equivalent performance (the null hypothesis is true), how frequently do the pairwise significance tests erroneously detect improvement in performance on large search and recommendation evaluation data?
    \item When one system outperforms another system (the null hypothesis is false), how powerful are the pairwise significance tests in detecting this improvement from large search and recommendation evaluation data?
\end{itemize}

We find that as the sample size increases, the false positive rates of sign and Wilcoxon tests increases. We also observe that as the sample size increases, the power of the tests increases, and the differences in the power amongst the tests diminish until there is no longer any distinguishable difference in their power. We find that Top-$N$ recommendation and large search experiments have a $100\%$ chance of finding statistically significant results even for very small effect sizes which may not be practically significant.

\section{Background and Related Work} \label{relatedwork}

The evaluation of information retrieval systems focuses on how well the system retrieves and ranks the retrieved documents. The Cranfield experiment is usually used for IR system evaluation \citep{cleverdon1967cranfield, voorhees2001philosophy}. The effectiveness of a system is measured over a set of queries or topics using set of relevance judgements $\textit{qrel}$.  Let $E_1, E_2, E_3, \dots, E_n$ be the effectiveness scores over $n$ topics of the experimental method $E$  and  $B_1, B_2, B_3, \dots, B_n$ be the effectiveness scores of the baseline method $B$. To compare their performance, we compute the difference of $B_i$ and $E_i$ such that $d_i = E_i-B_i$. 
We use the pairwise comparison statistical significance tests to determine if the mean of the observed difference $\bar{d_i}$ is significant or due to statistical random chance. The pairwise comparison statistical significance tests are commonly used when the data are paired, as when we evaluate two IR systems using the same topic sets \citep{sakai2018laboratory}.

Five pairwise tests and their null hypotheses are:
\begin{description}
\item[Student’s Paired $t$-test] The mean difference is zero \citep{fisher1936design}.
\item[Bootstrap shift test] $E$ and $B$ are samples from the same distribution \citep{noreen1989computer}.
\item[Randomization test] System $E$ and system $B$ have identically-distributed effectiveness scores \citep{box1978statistics}.
\item[Sign test] The median($d$)=$0$ \citep{wackerly2014mathematical}.
\item[Wilcoxon signed rank test] $d$ symmetric with median $0$ \citep{wackerly2014mathematical}.
\end{description}

The sign and Wilcoxon tests are used for IR experiments because they were recommended in the early account of statistical significance testing in IR \citep{rijsbergen1979information, zobel1998reliable}. Considering the mean or the median is inconsequential unless the data is skewed substantially \citep{ramachandran2020mathematical}. 

A number of studies have investigated the suitability of these statistical tests for the analysis of IR evaluation data particularly for the results of TREC-style experiments  \citep{hull1993using, sanderson2005information, zobel1998reliable, smucker2007comparison, parapar2020using, urbano2019statistical}. \citet{smucker2007comparison} used the randomization test as ground truth for comparing pairs of TREC runs and compared the other tests to it. They recommended the discontinuation of the Sign and Wilcoxon tests because they tend to disagree with the randomization test while the bootstrap and $t$-test agreed with it. \citet{urbano2019statistical} and \citet{parapar2020using} trained simulations to mimic evaluation results from historic TREC evaluations. Simulation allowed them to control whether the null hypothesis was true or not, and compute actual error rates and power. \citet{parapar2020using}'s method simulates new runs for the same topic while \citet{urbano2019statistical}'s approach simulates new topics for the same system. Based on their findings, \citet{urbano2019statistical} recommended the use of the $t$-test while \citet{parapar2020using} recommended the use of the sign and Wilcoxon tests.  There is some debate about these conflicting results~\citep{parapar2021testing}, whether they arise from \citeauthor{urbano2019statistical}'s use of parametric distributions, but \citet{urbano2021metric} argue that their methods would favor the Wilcoxon test if it was appropriate.

We treat the topic evaluation scores of a system as observations of a random variable
following \citet{sakai2018laboratory}. Therefore, we adopt the method proposed by \citet{urbano2018stochastic} for simulation.

\section{Data and Methods}
We used evaluation results from multiple tasks (passage ranking and top-$N$ recommendation) to study statistical test behavior. Each evaluation results dataset consists of a set of systems, a set of evaluation requests (a topic for search, and a user for recommendation), and effectiveness scores. We selected these evaluation datasets because the typify the datasets used in large search and top-$N$ recommendation experiments; the size of the evaluation requests ranges from $943$ to $ 32,509$ (substantially more than the $50$ evaluation requests typically used for the ad-hoc search TREC track), and the relevance data is sparse and highly incomplete.

Table \ref{tbl:eval_results} summarizes our evaluation results dataset.

\begin{table}[tbh]
    \caption{Summary of the evaluation results studied.}
    \label{tbl:eval_results}
    \centering
    \begin{tabular}{cccc}
        Dataset& Task & Systems & Requests\\
        \toprule
        \textsf{ML-100K} & Recommendation & 58 &943  \\
        \textsf{ML-25M} & Recommendation & 58 &32,509\\
        \textsf{AZ-Video} & Recommendation & 58 &6,000\\
        \textsf{MS MARCO} & Passage ranking & 63 &6,980 \\
    \end{tabular}
\end{table}

\subsection{Large-Scale Search}
We used a collection of MS-MARCO dev set runs provided by the MS-MARCO team \citep{bajaj2016ms}. We evaluated the runs with the official evaluation script they provided to participants\footnote{\url{https://github.com/microsoft/MSMARCO-Passage-Ranking/blob/master/ms_marco_eval.py}}.  MS-MARCO runs use the reciprocal rank ($\operatorname{RR}$) metric for effectiveness, as reported by the official evaluation script they provided to participants.

\subsection{Top-N Recommendation System}

We generated evaluation scores for top-$N$ recommendation by training four recommender system algorithms with different hyperparameters values (see Table \ref{tbl:algos}) using LensKit for Python (version 0.13) \citep{ekstrand2020lenskit}. The combinations of algorithms and associated hyperparameters values gave a total of $58$ systems.

We trained the algorithms on three publicly-available data sets (two from MovieLens \citep{harper2015movielens} --- \textsf{ML-100K} and \textsf{ML-25M} ---  and \textsf{AZ-Video} from Amazon \citep{he2016ups}). Table \ref{tbl:datasets} summarizes these datasets. 

\begin{table}[tbh]
    \caption{Summary of data sets.}
    \label{tbl:datasets}
    \centering
    \begin{tabular}{cccc}
        Dataset&\# ratings&\# requests & Density\\
        \toprule
        \textsf{ML-100K} & 100,000 & 1,000 & 6.3\% \\
        \textsf{ML-25M} & 25,000,000 & 162,000 & 0.26\%\\
        \textsf{AZ-Video} & 583,933 & 29,756 & 0.01\%
    \end{tabular}
\end{table}
 
For consistency with MS-MARCO, we used the $\operatorname{RR}$ metric to assess the quality of the recommendations generated for each user by the different trained models. We also computed $\operatorname{nDCG}$ scores but omit them for reasons of space. The evaluation of the generated recommendations resulted in a set of “runs” where a run represents a trained model and comprise of the evaluation scores for a set of users.

To train and evaluate an algorithm for a dataset, we split the dataset by partitioning the set of users into $5$ sets of test users. For each set of test users, we selected $20\%$ of a test user's interactions for testing and used the rest for training, along with data from the other users. For each algorithm, we:
\begin{itemize}
\item trained on the training data.
\item generated $100$ recommendations for each test user.
\item used the $\operatorname{RR}$ metrics to evaluate the generated recommendation lists. 
\end{itemize}

The selection of $20\%$ of a test user's rows implied that users with fewer than 5 ratings were not included in the test data. 

\begin{table}[tbh]
    \caption{Recommender algorithms used and their hyperparameter settings.}
    \label{tbl:algos}
    \centering\small
    \begin{tabular}{@{}ccl@{}}
    \textbf{Algorithm} & \textbf{Hyperparameter} & \textbf{Values}\\
     \toprule
     Item k-NN & nnbrs & $5, 10, 20, 30, 50$ \\ 
     \midrule
     User-based k-NN & nnbrs & $5, 10, 20, 30, 50$ \\ 
     \midrule
     \multirow{2}{*}{ALS BiasedMF} & als-features & $5, 10, 25, 50, 100, 200, 300, 500$\\ 
      & als-iterations & $5,10,20$ \\ 
     \midrule
     \multirow{2}{*}{ALS ImplicitMF} & als-features & $5, 10, 25, 50, 100, 200, 300, 500$ \\ 
       & als-iterations & $5,10,20$ \\ 
       \bottomrule
    \end{tabular}
\end{table}

\subsection{Experiments}
We designed the experiment to compare the false positive rates and power of the two-tailed statistical tests on the four datasets for sample sizes $n = 25, 50, 100$ to be consistent with sample sizes used in \citep{urbano2019statistical, parapar2020using} and $n=500, 1000, 5000, 10000, 20000$ to get closer to the sample sizes used in large search and recommendation experiments. We used the methodology proposed by \citet{urbano2018stochastic}, a generative stochastic simulation model which consists of two parts: (1) a system's effectiveness score distribution (marginal distribution for the system) and (2) a bivariate copula that models the dependence between pairs of runs (joint distribution). We simulated new effectiveness scores from the fitted model over arbitrarily topics.

To fit the model:
\begin{enumerate}
    \item Fit a marginal distribution $F_B$ to each run $B$ in the evaluation dataset such that $B \sim F_B$. We would refer to these as baseline runs. These marginal distributions are parametric and non-parametric which take the form of Truncated Normal, Beta, Beta-Binomial and discrete kernel smoothing distributions.
    \item Select a pair of runs ($B \sim F_B$, $E \sim F_E$) whose difference in mean scores are closest to an effect size $\delta$ and modify $F_E$ such that it is transformed with a new mean $\mu_E = \mu_B + \delta$. 
    \item Use the $\mathrm{CDF}$ of a baseline run $F_B$ to transform $B$ to pseudo-observations $U_B$ such that $U_B \sim$ Uniform(0, 1). 
    \item Fit the copula $C$ to every pair of $U_B$. 
\end{enumerate}

To measure the false positive (type I) error rate we simulated pairs of runs from the fitted model (in this scenario the null hypothesis is true) as follows:  
\begin{enumerate}
   \item Draw new pseudo-observations ($V_B$, $V_E$) from the fitted copula.
   \item Apply the inverse $\mathrm{CDF}$ of the marginal $B$ to $V_B$, $V_E$ to get the final effectiveness scores $B=F_B^{-1}(V_B)$ and $E=F_B^{-1}(V_E)$. Using the same marginal ensures that the effectiveness is the same between systems.
\end{enumerate}
 
We simulated $n$ new runs and computed the 2-tailed p-values for each of the statistical tests, repeating this process $10,000$ times for each combination of dataset, metric, and sample size for a total of $180,000$ trials. A statistically significant result by any of the test is counted as false positive against the test. 

To measure the power, we generated runs ($B$, $E$) with different effect sizes $\delta$ such that $\mu_E = \mu_B + \delta$ (in this scenario the null hypothesis is false). We generated the runs as follows:
\begin{enumerate}
    \item Select a fitted copula with baseline runs that have a difference in means close to $\delta$. The copula is associated with a baseline run $B$ and a transformed run $E$. 
    \item Draw pseudo-observations ($V_B$, $V_E$) from the selected copula.
    \item Apply the inverse $\mathrm{CDF}$ of $B$ to $V_B$, and the inverse $\mathrm{CDF}$ of $E$ to $V_E$ to get effective scores for new requests ($B = F_B^{-1}(V_B)$, $E = F_E^{-1}(V_E)$).    
\end{enumerate}

We simulated pairs of effectiveness scores of size $n$ with effect sizes $\delta$ and computed the 2-tailed p-values for each of the statistical tests. We repeated this process $2,500$ times for every combination of $\delta$ (in the range [0.01, 0.02, …, 0.1]), sample size $n$, metric, and dataset which gave us a total of $1,400,000$ trials. Under this model, the null hypothesis is false. Therefore, any statistically significant result by any of the test will count as true positive (power).

\section{Findings}

\begin{figure}[tb]
  \centering
  \includegraphics[width=\columnwidth]{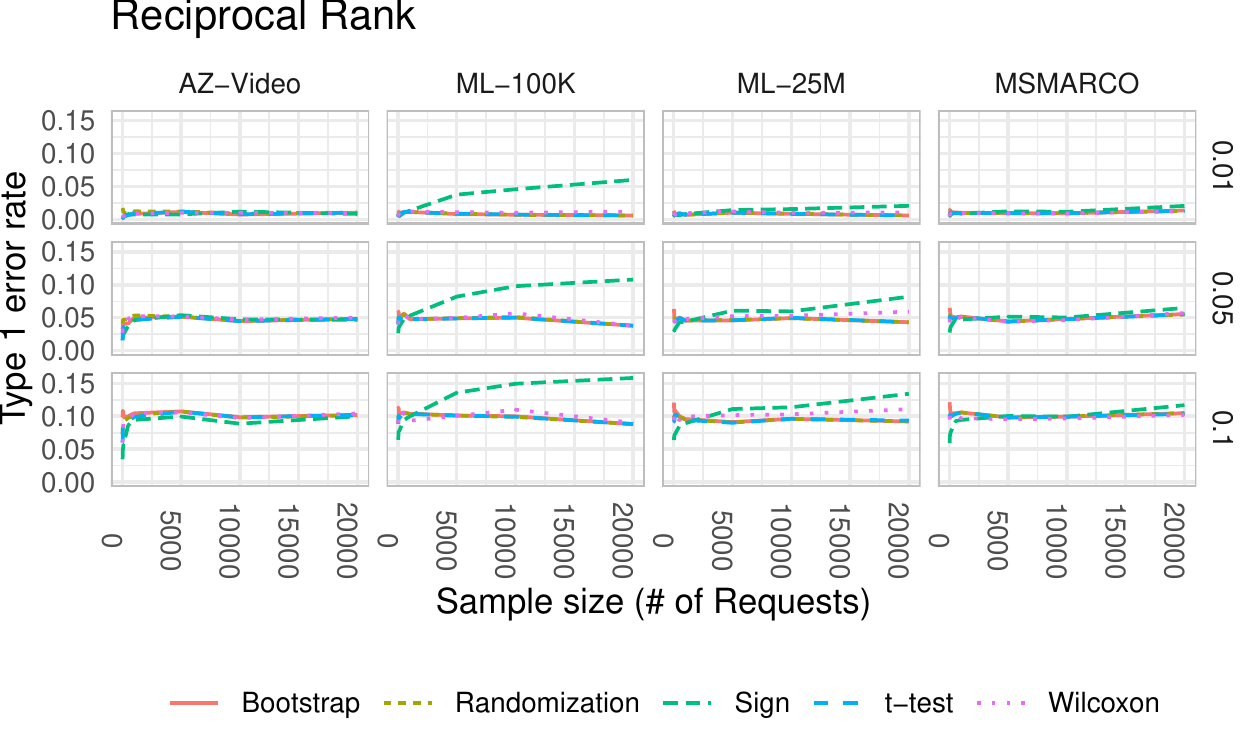}
  \caption{False positive rates of 2-tailed tests. Expected behavior: The false positive rate of a test should be directly proportional to a specified significance level ($\alpha = 0.01, 0.05, 0.1$).}
  \label{fig:type1_falsepositve}
\end{figure}
\begin{figure}[tbh]
  \centering
  \includegraphics[width=\columnwidth]{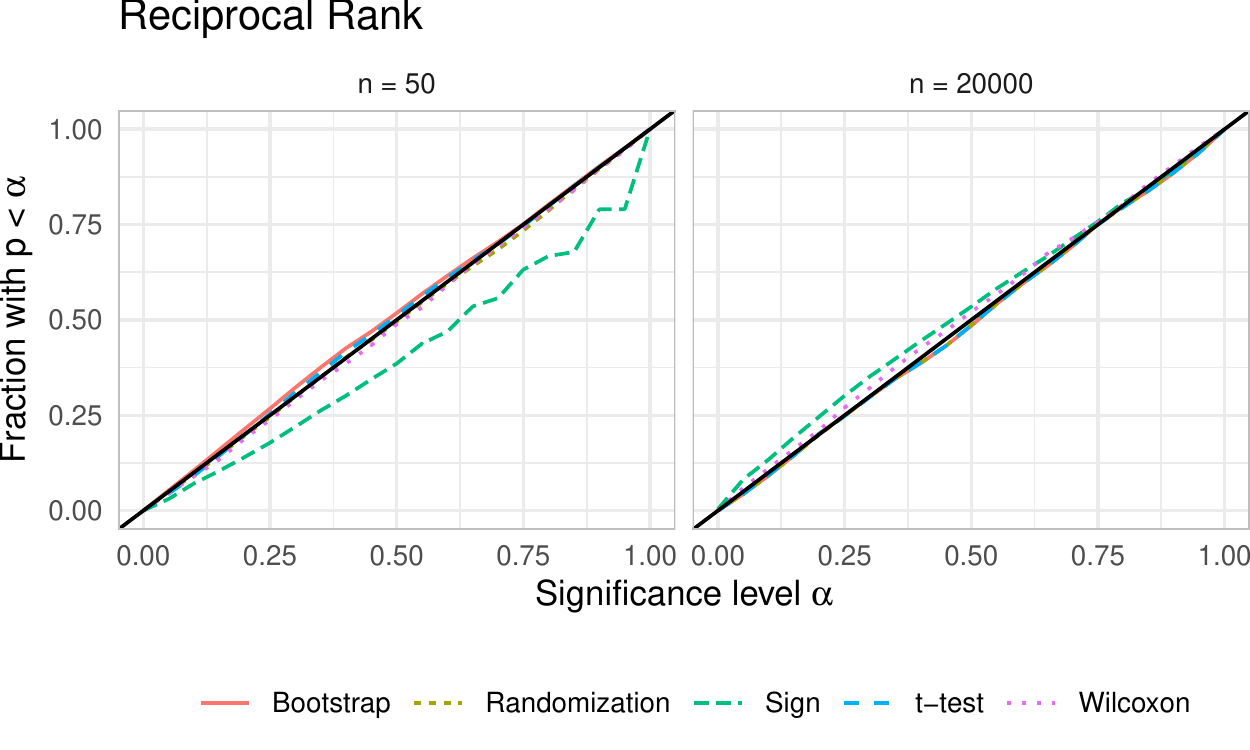}
  \caption{The calibration of the $p$-values of the statistical significance tests. A test is well calibrated if it is close to the diagonal. Fraction of mean difference with a $p$-value below $\alpha$ is approximately the same as $\alpha$.}
  \label{fig:type1_calibrated}
\end{figure}

Figures \ref{fig:type1_falsepositve} and \ref{fig:type1_calibrated} show the false positive error rates of the tests for different sample sizes at significance levels $\alpha = 0.01, 0.05, 0.1$ and the $p$-values calibration of the tests for a broader range of $\alpha$ respectively. When the null hypothesis is true (two systems have equivalent performance), a significance test is expected to be well calibrated (close to the diagonal) with false positive error rate that is equal to the significance level $\alpha$.  

From figure \ref{fig:type1_falsepositve}, we observe the following behavior of the significance tests on the different datasets:
\begin{itemize}
    \item For small sample sizes, the sign test makes less error than expected, the Wilcoxon test is close to the expected behavior, and the bootstrap makes more error than expected.
    \item As the sample size increases, the bootstrap approaches the expected behavior while the sign  and Wilcoxon test starts making more errors than expected. We observed a much more higher than expected error rates with the $\operatorname{nDCG}$ metric for the sign and Wilcoxon tests but omit them for reasons of space.
    \item The $t$-test and randomization tests are close to the expected behavior for all sample sizes.
\end{itemize}

Further, figure \ref{fig:type1_calibrated} shows that at $n=50$ the sign test has lower than expected false positives across a range of significance thresholds, while at $n=20000$, both sign and Wilcoxon have higher than expected error.   

\begin{figure}[tbh]
  \centering
  \includegraphics[width=\columnwidth]{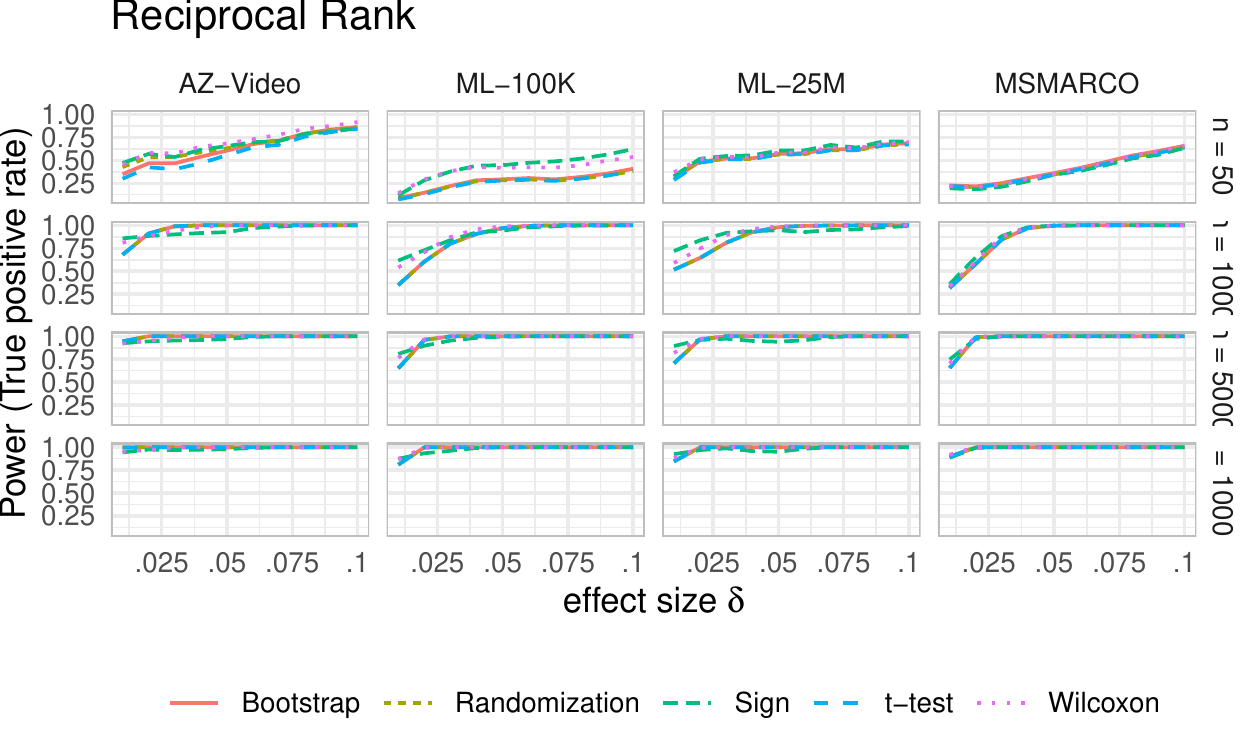}
  \caption{Power of statistical tests as a function of effect size and sample size at significance level: 0.05.}
  \label{fig:power_recip_rank100}
\end{figure}

Figure \ref{fig:power_recip_rank100} shows the power for the two-tailed tests at significance level $\alpha = 0.05$. We observe the following behavior of the significance tests on the different datasets:
\begin{itemize}
    \item  For small sample size ($n$=50), there are differences in the power of the tests. The sign and Wilcoxon tests are the most powerful and the $t$-test the least powerful for the $\operatorname{RR}$ metric.
    \item As the sample size increases the power of the tests increases, and the observed differences in their power decreases such that there is no distinguishable difference.
\end{itemize}

With large sample sizes, all tests readily find extremely small effects.

\section{Discussion and Conclusion}
Our empirical study of the false positive rates of the pairwise statistical tests with Top-$N$ recommendation and large search evaluation data, showed that as the sample size increases, the false positive rates of the sign and Wilcoxon tests increases. Under the null hypothesis, the difference distribution is expected to be symmetric about zero. Slight deviation from symmetry about zero is usually inconsequential and not statistically significant for small sample sizes. However, we observe that when the sample size is large, the sign and Wilcoxon tests find these slight deviations from perfect symmetry about zero in the difference distribution as statistically significant. This result is in agreement with findings from \citet{smucker2007comparison} and \citet{urbano2019statistical} that the Wilcoxon and sign tests are unreliable for the analysis of mean effectiveness. Our results demonstrates that this unreliability has even more serious consequences with the sample sizes found in large search and recommendation experiments. Therefore, we recommend that the sign and Wilcoxon tests should not be used for analyzing recommendation and large search evaluation data.

This study also analyzed the power of the pairwise significance tests in detecting performance improvement from large search and recommendation evaluation data. We found that for small sample size the power of the pairwise significance tests are distinguishable but as the sample size increases, the difference in the power amongst the tests begin to diminish until there is no longer any distinguishable difference in their power (see \ref{fig:power_recip_rank100}). While previous research \citep{urbano2019statistical, parapar2020using} made recommendations that were based on findings that some tests were more powerful than others, our result demonstrate that with Top-$N$ recommendation and large search evaluation data, most tests would have a $100\%$ chance of finding statistically significant results. Therefore, we can be confident that the effect size is precise and not actually $0$. However, we need to decide if the effect size is scientifically or practically meaningful. We recommend that both the $p$-value, and effect size be reported \citep{fuhr2018some, wasserstein2019moving}, and used for decision making.

Further research is needed that provide statistical techniques that provide meaningful results in light of the specific problems experienced in top-$N$ recommendation and large search evaluation.

\begin{acks}
This work partially supported by the National Science Foundation under Grant IIS 17-51278. Computation performed on Borah \citep{byrne2020borah}. MS-MARCO runs generously provided by Bhaskar Mitra and Nick Craswell.
\end{acks}

\balance
\bibliographystyle{ACM-Reference-Format}
\bibliography{bibfile}

\end{document}